\documentclass{elsartL}
\usepackage{graphicx}
\usepackage{amsmath}
\usepackage{amssymb}
\usepackage{mathrsfs}
\begin{document}

%\tikzstyle{decision} = [diamond, draw, fill=red!20, text width=7em, text badly centered, node distance=4cm, inner sep=0pt]
%\tikzstyle{block} = [rectangle, draw, fill=blue!20, text width=10em, text centered, rounded corners, minimum height=4em]
%\tikzstyle{line} = [draw, very thick, color=black!50, -latex']
%\tikzstyle{cloud} = [draw, ellipse,fill=yellow!20, node distance=2.5cm, minimum height=2em]

\begin{frontmatter}
\title{On using the Microsoft Kinect$^{\rm TM}$ sensors to determine the lengths of the arm and leg bones of a human subject in motion}
\author[IMS]{M.J.~Malinowski},
\author[IMS]{E.~Matsinos{$^*$}}
%\author[InIT]{R.~Jain},
%\author[InIT]{S.~Roth}
\address[IMS]{Institute of Mechatronic Systems, School of Engineering, Zurich University of Applied Sciences (ZHAW), Technikumstrasse 5, CH-8401 Winterthur, Switzerland}
%\address[InIT]{Institute of Applied Information Technology, School of Engineering, Zurich University of Applied Sciences (ZHAW), Steinberggasse 13, CH-8401 Winterthur, Switzerland}
%\address[IPT]{Institute of Physiotherapy, School of Health Professions, Zurich University of Applied Sciences (ZHAW), Technikumstrasse 71, CH-8401 Winterthur, Switzerland}
%\address[IMES]{Institute of Mechanical Systems, School of Engineering, Zurich University of Applied Sciences (ZHAW), Technikumstrasse 9, CH-8401 Winterthur, Switzerland}

\begin{abstract}The present study is part of a broader programme, exploring the possibility of involving the Microsoft Kinect$^{\rm TM}$ sensor in the analysis of human motion. We examine the output obtained from the two available 
versions of this sensor in relation to the variability of the estimates of the lengths of eight bones belonging to the subject's extremities: of the humerus (upper arm), ulna (lower arm, forearm), femur (upper leg), and tibia 
(lower leg, shank). Large systematic effects in the output of the two sensors have been observed.
\end{abstract}
{$^*$}{E-mail: evangelos[DOT]matsinos[AT]zhaw[DOT]ch, evangelos[DOT]matsinos[AT]sunrise[DOT]ch}
\end{frontmatter}

\section{\label{sec:Introduction}Introduction}

In a previous study \cite{mmr}, we established the theoretical background needed for the comparison of the output of measurement systems used in capturing data for the analysis of human motion. We developed our methodology for a 
direct application in case of the two versions of the Microsoft Kinect$^{\rm TM}$ (hereafter, simply `Kinect') sensor \cite{Kinect}, a low-cost, portable motion-sensing hardware device, developed by the Microsoft Corporation 
(Microsoft, USA) as an accessory to the Xbox $360$ video-game console (2010). The sensor is a webcamera-type, add-on peripheral device, enabling the operation of Xbox via gestures and spoken commands. The first upgrade of the 
sensor (`Kinect for Windows v2'), both hardware- and software-wise, tailored to the needs of Xbox One, became available for general development and use in July 2014. In Ref.~\cite{mm}, we applied our methodology \cite{mmr} to a 
comparative study of the two Kinect sensors and drew attention to significant differences in their output.

Previous attempts to validate the Kinect sensor for various medical/health-relating applications have been discussed in Ref.~\cite{mmr}. The present study is part of our research programme, investigating the possibility of 
involving (either of) the Kinect sensors in the analysis of the motion of subjects walking or running `in place', e.g., on commercially-available treadmills. If successful, Kinect could become an interesting alternative to 
marker-based systems (MBSs) in capturing data for motion analysis, one with an incontestably high benefit-to-cost ratio.

Studied herein is the dependence of the evaluated lengths of eight bones of the subject's extremities on the kinematical variables pertaining to the viewing of these bones by the sensor. The bones are: humerus (upper arm), ulna 
(lower arm, forearm), femur (upper leg), and tibia (lower leg, shank). The evaluated lengths of the subject's left and the right extremities are separately analysed~\footnote{The subject's left and right parts refer to what the 
subject perceives as the left and right parts of his/her body.}.

Ideally, the lengths of the bones of the subject's extremities should come out constant, irrespective of the orientation of these bones in space and of the viewing angle by the sensor. In reality, a departure from constancy is 
inevitable, given that the Kinect nodes represent centroids in the probability maps obtained from the captured data in each frame separately \cite{sh}; as such, the node-extraction process is subject to statistical fluctuations 
and is affected by different systematic effects in the three spatial directions. In the present work, we first examine the variability of the evaluated bone lengths with the variation of two angles describing the orientation of 
the bones in space, to be denoted hereafter as $\theta$ and $\phi$; $\theta$ is the angle of the bone with the vertical, whereas $\phi$ is the angle between the bone projection on the (anatomical) transverse plane and the $z$ 
axis (the direction associated with the depth in the images obtained with the sensors). Subsequently, we investigate the dependence of the evaluated lengths of these bones on the inclination angle with respect to Kinect's viewing 
direction. We pursue the determination of systematic effects in the Kinect-captured data, placing emphasis on establishing the similarities and the differences in the behaviour of the two sensors; in this respect, the present 
paper constitutes another comparative study of the two Kinect sensors, albeit from a perspective different to the one of Refs.~\cite{mmr,mm}.

The material in the present paper has been organised in four sections. In Section \ref{sec:Method}, we provide the details needed in the evaluation and in the analysis of the bone lengths. The results of the study are contained 
in Section \ref{sec:Results}. We discuss the implications of our findings in the last section.

\section{\label{sec:Method}On the evaluation and analysis of the bone lengths}

In the original sensor, the skeletal data (`stick figure') of the output comprises $20$ time series of three-dimensional (3D) vectors of spatial coordinates, i.e., estimates of the ($x$,$y$,$z$) coordinates of the $20$ nodes which 
the sensor associates with the axial and appendicular parts of the human skeleton. While walking or running, the subject faces the Kinect sensor. In coronal (frontal) view of the subject (sensor view), the Kinect coordinate system 
is defined with the $x$ axis (medial-lateral) pointing to the left (i.e., to the right part of the body of the subject being viewed), the $y$ axis (vertical) upwards, and the $z$ axis (anterior-posterior) away from the sensor, see 
Fig.~\ref{fig:CS}. In the upgraded sensor, five new nodes have been appended at the end of the list: one is a body node, whereas the remaining nodes pertain to the subject's hands. In both versions, parallel to the video image, 
Kinect captures an infrared image, which enables the extraction of information on the depth $z$. The sampling rate in the Kinect output (for the video and the skeletal data, in both versions of the sensor) is $30$ Hz. As the Kinect 
output has already been detailed twice, in Sections 2.1 of Refs.~\cite{mmr,mm}, there is no need to further describe it here.

The `upper' endpoint (upper or proximal extremity) of each bone will be identified by the subscript $S$, the `lower' endpoint (lower or distal extremity) by $E$. The upper and lower endpoints of the bones refer to the upright 
(erect) standing (rest) position of the subject (standard anatomical position); in this position, $y_S > y_E$ in all cases. The four bones pertaining to the upper extremities (arms) are defined on the basis of the Kinect nodes 
SHOULDER\_LEFT, ELBOW\_LEFT, and WRIST\_LEFT (left side); SHOULDER\_RIGHT, ELBOW\_RIGHT, and WRIST\_RIGHT (right side). The four bones pertaining to the lower extremities (legs) are defined using the Kinect nodes HIP\_LEFT, 
KNEE\_LEFT, and ANKLE\_LEFT (left side); HIP\_RIGHT, KNEE\_RIGHT, and ANKLE\_RIGHT (right side).

Denoting the bone length as $L=\sqrt{(x_S-x_E)^2+(y_S-y_E)^2+(z_S-z_E)^2}$, the angle $\theta$ is estimated via the expression
\begin{equation} \label{eq:EQ01}
y_S - y_E \equiv \Delta y = L \, \cos\theta \, \, \, .
\end{equation}
Evidently, $\theta \in [0,\pi]$; in the standard anatomical position, $\theta=0^\circ$ (ideally). Additionally,
\begin{equation} \label{eq:EQ02}
x_S - x_E \equiv \Delta x = L \, \sin\theta \, \sin\phi
\end{equation}
and
\begin{equation} \label{eq:EQ03}
z_S - z_E \equiv \Delta z = L \, \sin\theta \, \cos\phi \, \, \, .
\end{equation}
Obviously, $\phi=0^\circ$ when the bone projection on the transverse plane ($x$,$z$) is antiparallel to the $z$ axis (i.e., when the projection of the vector $\overrightarrow{SE}$ on that plane points in the direction of the 
sensor). Denoting the projected bone length on the transverse plane as $L_{xz}=\sqrt{(\Delta x)^2+(\Delta z)^2}=L \, \sin\theta$, we obtain the relation:
\begin{equation} \label{eq:EQ04}
\cos\phi = \frac{\Delta z}{L_{xz}}
\end{equation}
for all $\theta$ values, exempting $\theta=0$ and $\theta=\pi$ where the projected length $L_{xz}$ vanishes; although cases in the data where $\theta$ comes out identical to $0$ or $\pi$ have not been found, we will assign (for 
the sake of mathematical completeness) the value of $0$ to $\phi$ in both cases. Evidently, $\phi \in [0,2\pi)$. In the present study, we will assume left/right symmetry (in the description of the motion of the bones) and restrict 
the angle $\phi$ in the $[0,\pi]$ domain.

Of relevance in the context of the present work is the angle at which the Kinect sensor views a bone. The inclination angle is obtained from the data as follows. Referring to Fig.~\ref{fig:CS}, we define the position vectors 
pertaining to the bone endpoints as $\vec{r}_S=(x_S,y_S,z_S)$ and $\vec{r}_E=(x_E,y_E,z_E)$. The unit vector, normal to the KSE plane, is obtained via the expression:
\begin{equation} \label{eq:EQ05}
\hat{u}_N = \frac{\vec{r}_E \times \vec{r}_S}{\lVert \vec{r}_E \times \vec{r}_S \rVert} \, \, \, .
\end{equation}
(We have not found instances in the data involving parallel or antiparallel vectors $\vec{r}_S$ and $\vec{r}_E$.) The position vector of the midpoint M of the SE segment is given by
\begin{equation} \label{eq:EQ06}
\vec{r}_M = \frac{\vec{r}_S+\vec{r}_E}{2}
\end{equation}
and the unit vector along that direction by
\begin{equation} \label{eq:EQ06_1}
\hat{u}_R = \frac{\vec{r}_M}{\lVert \vec{r}_M \rVert} \, \, \, .
\end{equation}

The unit vector $\hat{u}$, on the KSE plane, orthogonal to $\hat{u}_R$, is obtained by the expression
\begin{equation} \label{eq:EQ06_2}
\hat{u} = \frac{\hat{u}_R \times \hat{u}_N}{\lVert \hat{u}_R \times \hat{u}_N \rVert} = \hat{u}_R \times \hat{u}_N \, \, \, .
\end{equation}
The inclination $\omega$ is the angle between $\overrightarrow{SE} \equiv \vec{r}_E - \vec{r}_S$ and $\hat{u}$; evidently,
\begin{equation} \label{eq:EQ07}
\cos\omega = \frac{(\vec{r}_E - \vec{r}_S) \cdot \hat{u}}{\lVert \vec{r}_E - \vec{r}_S \rVert} \, \, \, .
\end{equation}
As we are interested in the inclination of each bone, not in its orientation with respect to the unit vector $\hat{u}$, we will use $\lvert \cos\omega \rvert$ as free variable. It is expected that the reliability of the 
Kinect-evaluated bone lengths should increase with increasing $\lvert \cos\omega \rvert$.

\section{\label{sec:Results}Results}

The data acquisition involved one male adult (ZHAW employee), with no known motion problems, walking and running on a commercially-available treadmill (Horizon Laufband Adventure 5 Plus, Johnson Health Tech.~GmbH, Germany). The 
placement of the treadmill in the laboratory of the Institute of Mechatronic Systems (School of Engineering, ZHAW), where the data acquisition took place, was such that the subject's motion be neither hindered nor influenced in 
any way by close-by objects. Prior to the data-acquisition sessions, the Kinect sensors were properly centred and aligned. The sensors were then left at the same position, untouched throughout the data acquisition.

It is worth mentioning that, as we are interested in capturing the motion of the subject's lower-leg parts (i.e., of the ankle and foot nodes), the Kinect sensors must be placed at such a height that the number of lost lower-leg 
signals be kept reasonably small. Our past experience dictated that the Kinect sensor should be placed close to the minimal height recommended by the manufacturer, namely around $2$ ft off the (treadmill-belt) floor. Placing the 
sensor higher (e.g., around the midpoint of the recommended range of values, namely at $4$ ft off the treadmill-belt floor) leads to many lost lower-leg signals (the ankle and foot nodes are not properly tracked), as the lower 
leg is not visible by the sensor during a sizeable fraction of the gait cycle, shortly after toe-off.

The Kinect sensor may lose track of the lower parts of the subject's extremities (wrists, hands, ankles, and feet) for two reasons: either due to the particularity of the motion of the extremity in relation to the position of the 
sensor (e.g., the identification of the elbows, wrists, and hands becomes problematic in some postures, where the Kinect viewing angle of the ulnar bone is small) or due to the obstruction of the extremities of the human body 
(behind the subject) for a fraction of the gait cycle. Assuming that these instances remain rare (e.g., below about $3 \%$ of the available data in each time series, i.e., no more than one lost frame in $30$), the missing values 
may be reliably obtained (interpolated) from the well-determined (tracked) data. Although, when normalised to the total number of available values, the untracked signals usually appear `harmless', particular attention was paid 
in order to ensure that no node be significantly affected.

Five velocities were used in the data acquisition: walking-motion data were acquired at $5$ km/h; running-motion data at $8$, $9$, $10$, and $11$ km/h. At each velocity setting, the subject was given $1$ min to adjust his movements 
comfortably to the velocity of the treadmill belt. The Kinect output spanned slightly over $2$ min at each velocity. The variation of the distance between the subject and the Kinect sensors was monitored during the data acquisition; 
it ranged between about $2.5$ and $2.9$ m, well within the limits of use of the sensors set by the manufacturer. The recording on the two measurement systems started simultaneously.

At each sampled frame, the bone lengths were calculated from the Kinect output and were histogrammed in ($\cos\theta$, $\cos\phi$) cells; the same values were also histogrammed in bins of $\lvert \cos\omega \rvert$. Twenty bins 
per angular direction were used in the former case, forty in the latter. Averages, as well as the standard errors of the means, were calculated in all cells/bins containing at least ten entries; cells/bins with fewer entries were 
ignored.

Techniques yielding accurate results for in-vivo measurements of the human long bones include conventional (planar) radiography, CT scanning (e.g., see Ref.~\cite{skrkhv}), Raman spectroscopy \cite{mdgcrp}, and ultrasonic scanning 
\cite{lcltzn}. Unwilling or unable to use any of these techniques, we obtained static measurements of the subject's bone lengths with an MBS (AICON 3D Systems GmbH, Germany) \cite{aicon}. Our system features two digital cameras, a 
control unit, and a high-end personal computer (where the visualisation software is installed). The MoveInspect Technology HF$\mid$HR reconstructs the 3D coordinates of the centres of markers (adhesive targets, reflective balls) 
which are simultaneously viewed by the two cameras; the typical uncertainty in the determination of the 3D coordinates of each marker centre is below $100$ $\mu$m, i.e., negligible when compared to other uncertainties, namely to 
those pertaining to the placement of the markers and to skin motion. The bone lengths, not to be identified herein with the suprema of the distances of any two points belonging to the bones regarded as 3D objects, were defined as 
follows.
\begin{itemize}
\item Humerus: from the centre of the humeral greater tubercle (`coinciding' with the Kinect shoulder node) to the humeral lateral epicondyle. As it is not straightforward to identify the centre of the humeral greater tubercles, 
flat markers were placed on the greater tuberosities and the resulting humeral-bone lengths were reduced~\footnote{The description of the algorithm, used in the determination of the 3D positions of the skeletal joints of the subject 
being viewed with Kinect, may be found in Ref.~\cite{sh}. The `3D positions of the joints' of Ref.~\cite{sh} are essentially produced from the `3D positions of the projections of the joints onto the front part of the subject's body' 
after applying a `shift' in depth (i.e., from the surface to the interior of the subject's body), namely a constant offset ($\zeta_c$) of $39$ mm (see end of Section 3 of Ref.~\cite{sh}). We will assume that the same correction is 
also applied to the infrared image in the vertical ($y$) direction (and, albeit not relevant herein, also in the medial-lateral ($x$) direction), in order to yield the $y$ (and $x$) coordinates of the shoulder joints. Therefore, 
when comparing it with the Kinect-evaluated humeral-bone length, the MBS-evaluated length will be reduced by $\zeta_c$.} by $39$ mm.
\item Ulna: from just below the humeral medial epicondyle, on top of the humeral trochlea, to the ulnar styloid process.
\item Femur: from the centre of the femoral head to just below the femoral medial condyle, on top of the medial meniscus. As the identification of the centres of the femoral heads, using a non-invasive anthropometric technique, 
is neither easy nor accurate, we obtained the hip positions by following the indirect approach of Ref.~\cite{dotg}, featuring four flat markers on the surface of the subject's body (see Subsection 2.2 of Ref.~\cite{mmr}). The 
knee positions were identified in two ways: either using $\varnothing14$-mm reflective balls on top of the medial menisci and applying a correction (lateral shift) similar to the one applied in Ref.~\cite{dotg} or using plat 
markers on the patellae (knee caps) and applying a correction in depth (half of the width of the subject's leg, at the knee level, in sagittal view); the two results were found almost identical.
\item Tibia: from just below the femoral medial condyle, on top of the medial meniscus, to the medial malleolus.
\end{itemize}
All measurements were acquired in the upright position. The extracted values of the subject's bone lengths are listed in Table \ref{tab:aicon}.

The average values, along with the rms (root-mean-square) values of the corresponding distributions, of the Kinect-evaluated bone lengths are shown in Table \ref{tab:Lengths}, separately for the two Kinect sensors. A noticeable 
difference between the two sets of values occurs in the femoral-bone lengths, the values of which come out, in all cases, significantly smaller when using the data of the upgraded sensor; this underestimation is equivalent to an 
effect between $1.69$ and $2.42$ standard deviations in the normal distribution. The quoted uncertainties in Table \ref{tab:Lengths} are large, indicating that systematic effects come into play, thus suggesting further analysis 
of the bone lengths, in terms of the kinematical variables $\theta$, $\phi$, and $\omega$.

\subsection{\label{sec:Results1}Profile scatter plots of the bone lengths in ($\cos\theta$,$\cos\phi$) cells}

Walking and running motions have different signatures; the differences are both quantitative and qualitative: the ranges of motion in running are larger, generally expected to increase with increasing velocity; qualitatively, the 
walking motion is characterised by extended elbow joints, the running motion by flexed ones. This last dissimilarity affects the detection of the elbows and of the wrists at several postures in running motion, inevitably 
introducing systematic effects in the evaluation of the humeral- and ulnar-bone lengths.

Our first study of the systematic effects in the evaluation of the bone lengths involved profile scatter plots in ($\cos\theta$,$\cos\phi$) cells. For convenience, the following definitions will be used in the short description 
of the results obtained in this part of the analysis.
\begin{itemize}
\item Forward (ventral) placement of a bone corresponds to a position of its lower-endpoint joint more proximal to the Kinect sensor than the mid-coronal plane; `very forward' placement refers to $\cos\phi>0.9$.
\item Backward (dorsal) placement of a bone corresponds to a position of its lower-endpoint joint more distal to the Kinect sensor than the mid-coronal plane; `very backward' placement refers to $\cos\phi<-0.9$.
\end{itemize}

In general, the walking motion is restricted to small $\theta$ values; only in the case of the ulna does the motion extend to $\theta \approx 60^\circ$. The humeral-bone lengths came out $\cos\phi$-dependent; systematically 
smaller values were extracted in forward placement, larger in backward. Regarding the ulna, the $\theta$ domain is somewhat enlarged in very forward placement, when the elbow joint is (usually) flexed. Regarding the tibia, a 
significant dependence of the evaluated length on $\cos\phi$ was seen in very backward placement; the effect was maximal around the most distal position of the ankle (with respect to the Kinect sensor), where the shank is not 
viewed sufficiently well by the sensor.

Compared to walking motion, the domain of the $\theta$ values is significantly enhanced in running (as expected). The evaluated humeral-bone lengths were (again) found $\cos\phi$-dependent. Regarding the ulna, the $\cos\phi$ 
values were large (restriction of the motion to very forward placement). For the tibial bones, the evaluated lengths in very backward placement were found to be significantly larger than those obtained elsewhere. Again, the 
largest bone lengths were extracted for large $\theta$ and $\phi$ values (large dorsal elevation of the lower leg), where the tibial bone is not viewed sufficiently well by the Kinect sensor.

In all cases, the relative minimax variation $d_r = 2 (M-m) / (M+m)$ (where $M$ and $m$ stand for corresponding maximal and minimal values, respectively) of the extracted values of the bone lengths was large. The $d_r$ ranges 
(over all velocities) were: for the humerus, $3$-$22 \%$ (original sensor) and $4$-$20 \%$ (upgraded sensor); for the ulna, $7$-$29 \%$ (original sensor) and $8$-$35 \%$ (upgraded sensor); for the femur, $4$-$13 \%$ (original 
sensor) and $7$-$17 \%$ (upgraded sensor); and for the tibia, $18$-$24 \%$ (original sensor) and $24$-$26 \%$ (upgraded sensor). Overall, the matching of the results between the two sensors in the case of running motion was not 
satisfactory; the values of Pearson's correlation coefficient (on the $d_r$ values of the eight bone lengths at fixed velocity) showed a pronounced velocity dependence, ranging from $0.958$ at $5$ km/h to $0.260$ at $11$ km/h.

\subsection{\label{sec:Results2}Profile histograms of the bone lengths in $\lvert \cos\omega \rvert$ bins}

The analysis described in the previous subsection is suggestive of the direction which the investigation must next turn to. In short, it appears plausible to assume that the goodness of the evaluation of the bone lengths does not 
depend separately on the angles $\theta$ and $\phi$, but on another quantity, namely on the viewing angle of the particular bone by the Kinect sensor. It is reasonable to expect that the accuracy in the evaluation of each bone 
length depends on the inclination of that bone with respect to the Kinect viewing direction; when a bone is viewed by the Kinect sensor at almost right angle, its length should be evaluated more reliably. The appropriate free 
variable in this investigation, $\lvert \cos\omega \rvert$, is obtained via Eq.~(\ref{eq:EQ07}).

\begin{figure}
\begin{center}
\includegraphics [width=15.5cm] {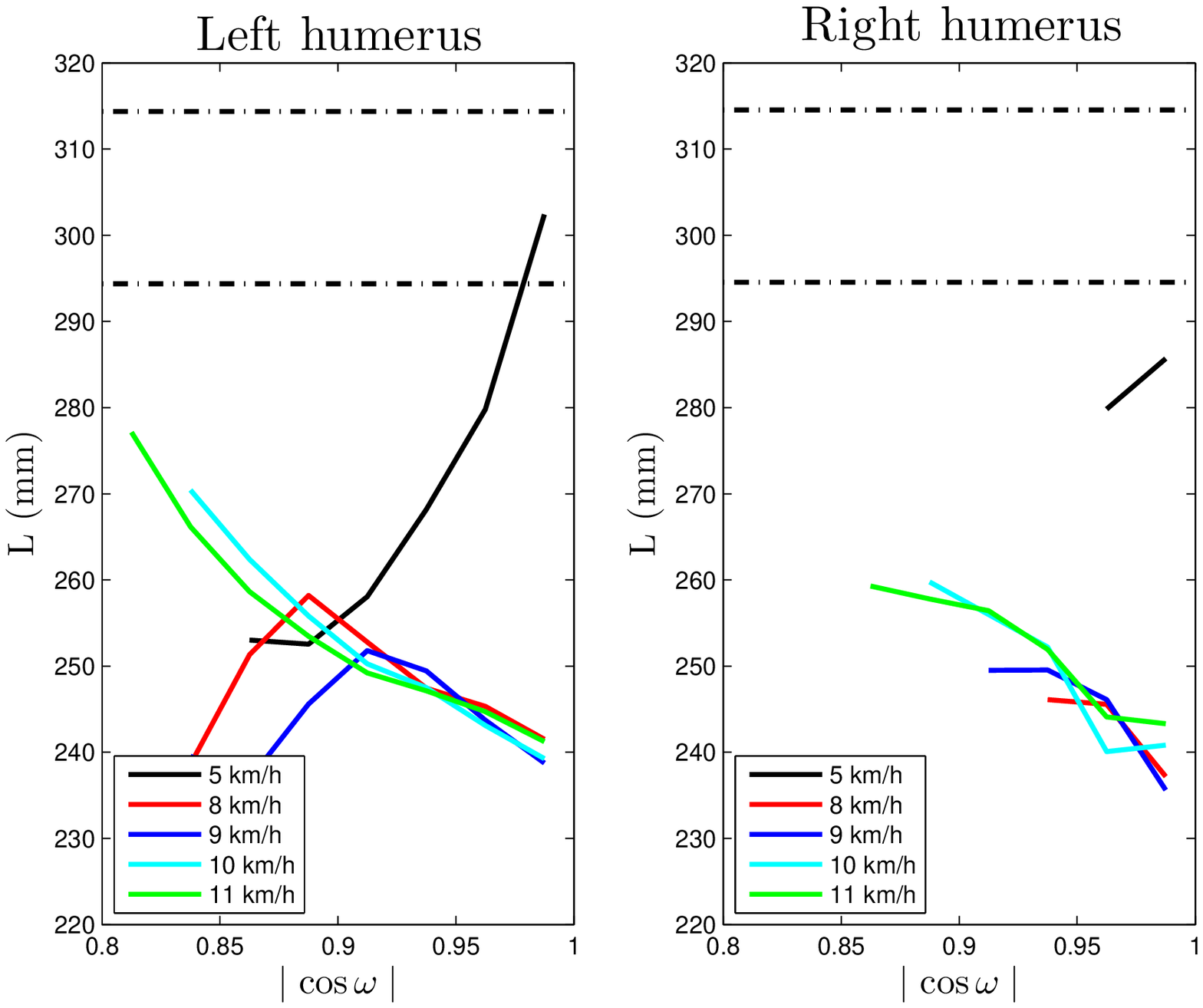}
%\vspace{-6cm}
\caption{\label{fig:LengthHumerusProfV1}The profile histogram of the humeral-bone length (separately for the left- and right-side bones) in $\lvert \cos\omega \rvert$ bins; the data has been captured with the original Kinect 
sensor. In walking motion, the swinging of the right arm of the subject used in our data acquisition, is very small.}
\end{center}
\end{figure}

\begin{figure}
\begin{center}
\includegraphics [width=15.5cm] {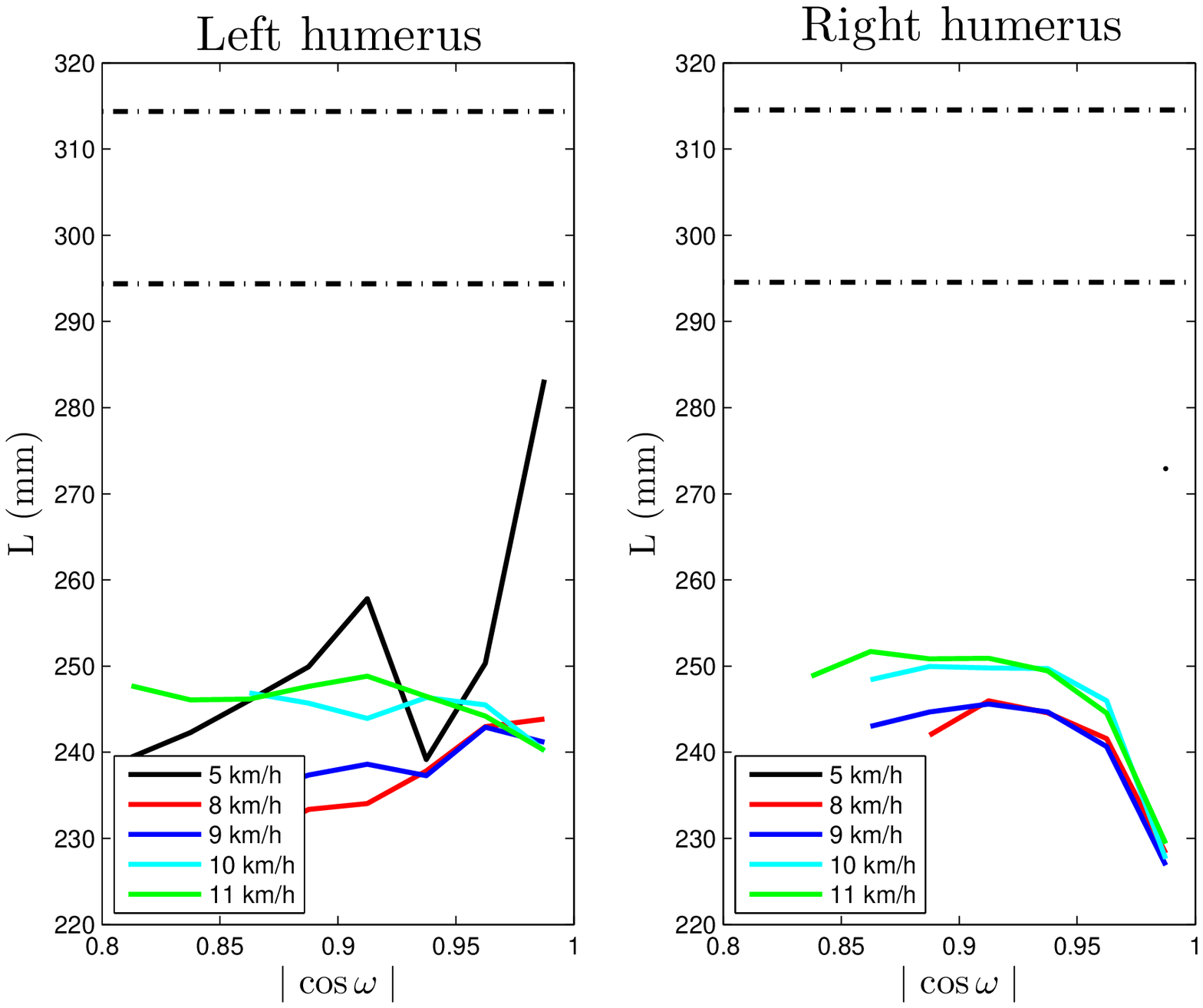}
%\vspace{-6cm}
\caption{\label{fig:LengthHumerusProfV2}The profile histogram of the humeral-bone length (separately for the left- and right-side bones) in $\lvert \cos\omega \rvert$ bins; the data has been captured with the upgraded Kinect 
sensor. In walking motion, the swinging of the right arm of the subject used in our data acquisition, is very small.}
\end{center}
\end{figure}

\begin{figure}
\begin{center}
\includegraphics [width=15.5cm] {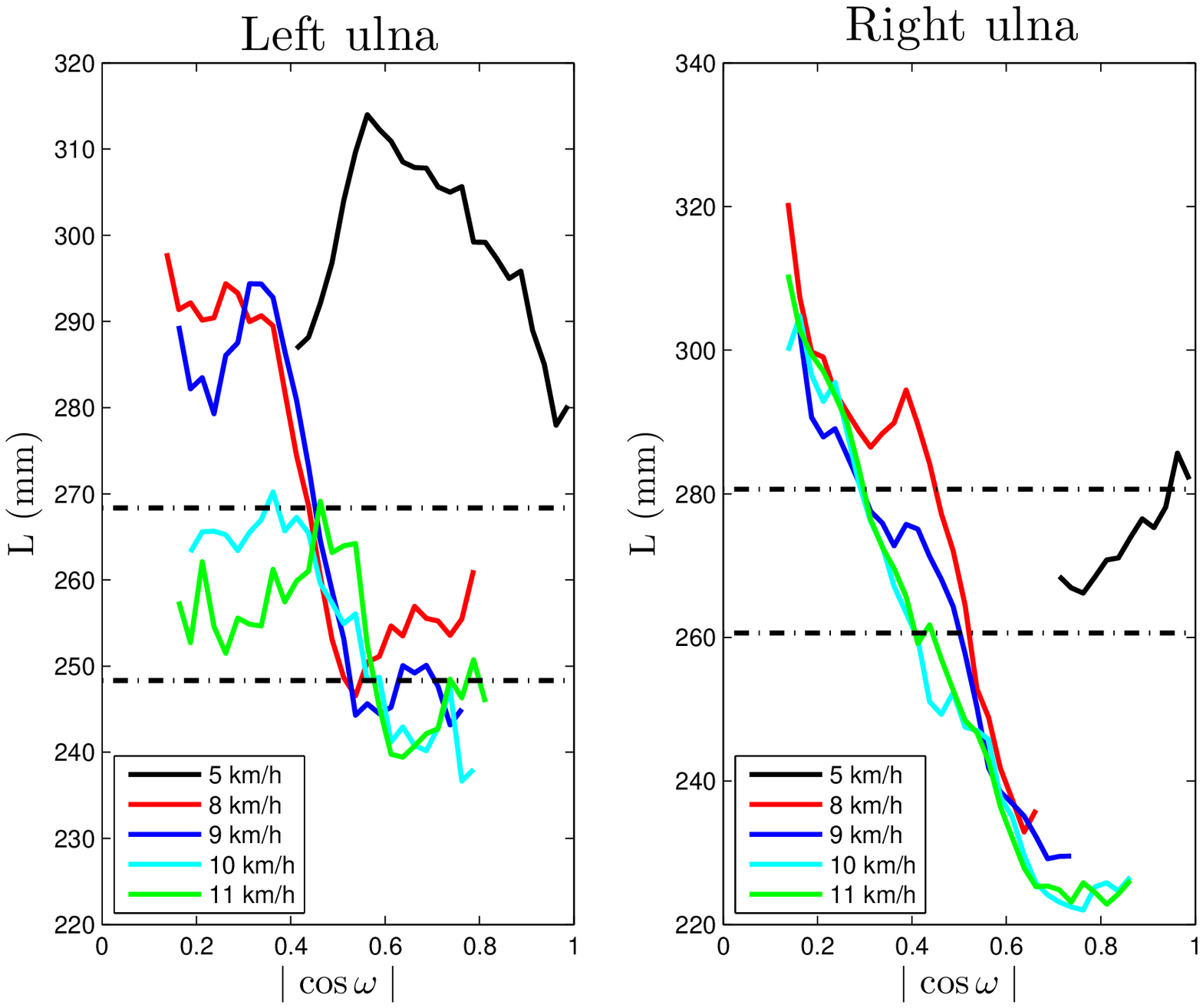}
%\vspace{-6cm}
\caption{\label{fig:LengthUlnaProfV1}The profile histogram of the ulnar-bone length (separately for the left- and right-side bones) in $\lvert \cos\omega \rvert$ bins; the data has been captured with the original Kinect sensor. 
In walking motion, the swinging of the right arm of the subject used in our data acquisition, is very small.}
\end{center}
\end{figure}

\begin{figure}
\begin{center}
\includegraphics [width=15.5cm] {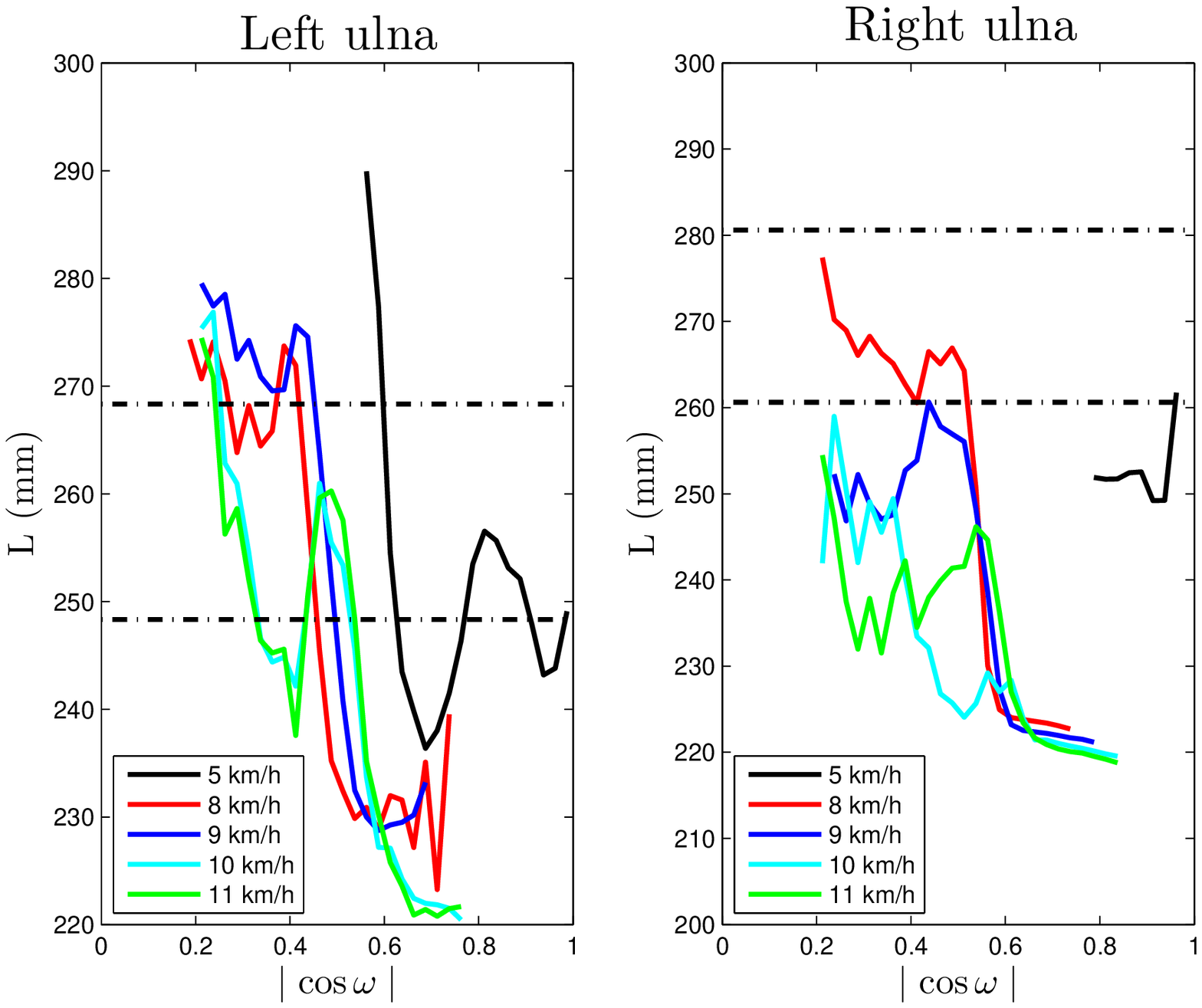}
%\vspace{-6cm}
\caption{\label{fig:LengthUlnaProfV2}The profile histogram of the ulnar-bone length (separately for the left- and right-side bones) in $\lvert \cos\omega \rvert$ bins; the data has been captured with the upgraded Kinect sensor. 
In walking motion, the swinging of the right arm of the subject used in our data acquisition, is very small.}
\end{center}
\end{figure}

\begin{figure}
\begin{center}
\includegraphics [width=15.5cm] {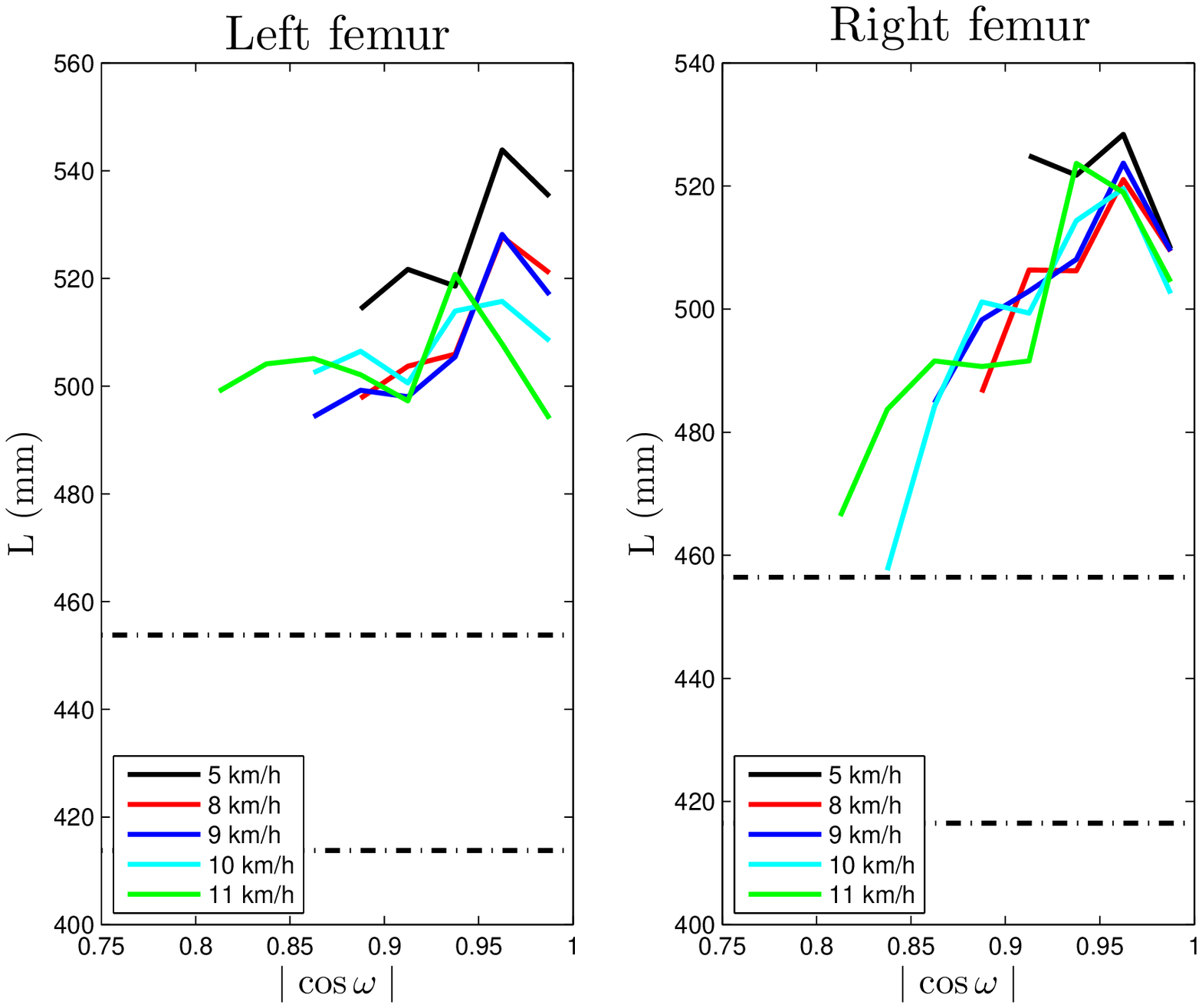}
%\vspace{-6cm}
\caption{\label{fig:LengthFemurProfV1}The profile histogram of the femoral-bone length (separately for the left- and right-side bones) in $\lvert \cos\omega \rvert$ bins; the data has been captured with the original Kinect sensor.}
\end{center}
\end{figure}

\begin{figure}
\begin{center}
\includegraphics [width=15.5cm] {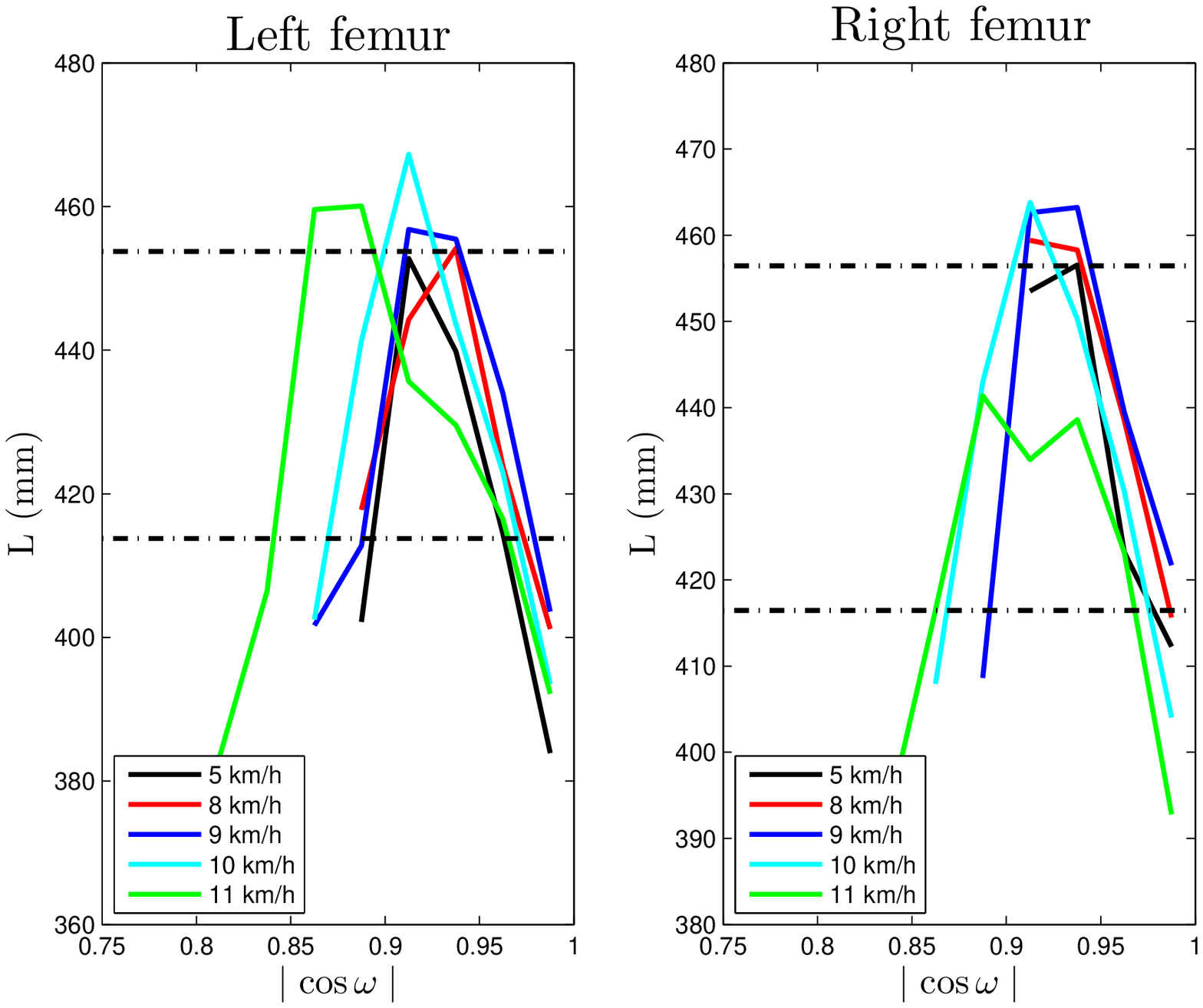}
%\vspace{-6cm}
\caption{\label{fig:LengthFemurProfV2}The profile histogram of the femoral-bone length (separately for the left- and right-side bones) in $\lvert \cos\omega \rvert$ bins; the data has been captured with the upgraded Kinect sensor.}
\end{center}
\end{figure}

The results, obtained from the data analysis, are shown in Figs.~\ref{fig:LengthHumerusProfV1}, \ref{fig:LengthUlnaProfV1}, \ref{fig:LengthFemurProfV1}, and \ref{fig:LengthTibiaProfV1} for the original sensor; in 
Figs.~\ref{fig:LengthHumerusProfV2}, \ref{fig:LengthUlnaProfV2}, \ref{fig:LengthFemurProfV2}, and \ref{fig:LengthTibiaProfV2} for the upgraded sensor. We now discuss these plots.

\begin{itemize}
\item {\bf Humerus}. There can be little doubt that both Kinect sensors systematically underestimate the humeral-bone length regardless of the type of the motion (walking or running). Only in the case $\lvert \cos\omega \rvert \to 1$ 
in the walking-motion data are the estimates close to the results of the static measurements. All humeral-bone lengths obtained for $\lvert \cos\omega \rvert \to 1$ in the case of the running-motion data agree well, and are about 
$65$-$75$ mm short of the results of the static measurements. This discrepancy is due to the misplacement of the arm joints during running, in particular of the elbows, which are kept in flexed position throughout the gait cycle. 
Compared to the original sensor, the upgraded sensor generally yields slightly shorter humeral-bone lengths.
\item {\bf Ulna}. In the data obtained with the original sensor, some left/right asymmetry in the results is visible. The range of variation of the ulnar-bone length was found maximal (about $10$ cm) in the case of the original 
sensor, for the right ulna. Regarding the walking-motion data for $\lvert \cos\omega \rvert \to 1$, the extracted values do not disagree with the results of the static measurements.
\item {\bf Femur}. The femoral-bone length, obtained with the original sensor, is seriously overestimated in all cases. The hip positions are mostly responsible for this discrepancy. (In Ref.~\cite{mm}, we reported that the 
waveforms for the hips, obtained from the two versions of the sensor, do not match well.)
\item {\bf Tibia}. In all cases, both sensors seriously overestimate the tibial-bone lengths for small viewing angles and underestimate them for $\lvert \cos\omega \rvert \to 1$. The range of variation of the tibial-bone length 
was found sizeable, namely between $9$ and $11$ cm. On the other hand, a nearly monotonic behaviour is observed in Figs.~\ref{fig:LengthTibiaProfV1} and \ref{fig:LengthTibiaProfV2}, and the results from the walking- and the 
running-motion data generally appear to be consistent between the two sensors.
\end{itemize}

It must be mentioned that the dependence of the bone lengths $L$ on $\lvert \cos\omega \rvert$ is expected to be monotonic. Visual inspection of Figs.~\ref{fig:LengthHumerusProfV1}-\ref{fig:LengthTibiaProfV2} reveals that this is 
not always the case. One reason for the observed departure from a monotonic behaviour (yet not the only one) is that the inclination $\omega$ is estimated from the Kinect-captured data; systematic effects also affect this estimation.

The results, reported in the present section, have also been checked against influences from `cross-talk', which might be present in the output of the two Kinect sensors. Both sensors use reflected infrared light in order to 
yield information on the depth in the captured images; one might thus argue that, in case of a simultaneous data acquisition, they distort one another's recording. To clarify this issue, data (using the same subject and velocity 
settings) were acquired with the two measurement systems `serially', first with the original sensor, subsequently with the upgraded one; in both cases, the second sensor was switched off (but was not removed from the mount). The 
differences to the results reported herein were inessential. Our conclusion is that the aforementioned discrepancies cannot be due to an interaction between the two sensors.

\section{\label{sec:Conclusions}Conclusions}

The present paper addressed one use of the output of the two Microsoft Kinect$^{\rm TM}$ (hereafter, simply `Kinect') sensors, namely the evaluation of the lengths of eight bones pertaining to the extremities of one subject walking 
and running on a treadmill: of the humerus (upper arm), ulna (lower arm, forearm), femur (upper leg), and tibia (lower leg, shank). For comparison, static measurements of these bone lengths were obtained with a marker-based system.

The evaluated lengths of the left and right parts of the subject's extremities have been separately analysed. The constancy of the lengths of these eight bones in terms of the variation of the two angles involved in the viewing, 
$\theta$ of Eq.~(\ref{eq:EQ01}) and $\phi$ of Eq.~(\ref{eq:EQ04}), was examined. We have also investigated the dependence of the bone lengths on the inclination angle $\omega$ of Eq.~(\ref{eq:EQ07}) with respect to Kinect's 
viewing direction. We pursued the analysis of systematic effects in the output, emphasising on the similarities and the differences between the two sensors; in this respect, the present study is another comparative study of the 
two Kinect sensors, albeit from a perspective different to that of Refs.~\cite{mmr,mm}.

The walking motion is characterised by extended elbow joints and small $\theta$ values in the case of the humerus and of the femur; in the case of the ulna, the $\theta$ values reached about $60^\circ$. The running motion is 
characterised by flexed elbow joints and (compared to the walking motion) extends more in $\theta$. The motion of the ulnar bones is different in walking and running; in the latter case, the ulnar motion extends in $\theta$, 
rather than in $\phi$. A major overestimation of the tibial-bone lengths occurs in very backward placement, where these bones cannot be viewed sufficiently well by the Kinect sensors.

\begin{figure}
\begin{center}
\includegraphics [width=15.5cm] {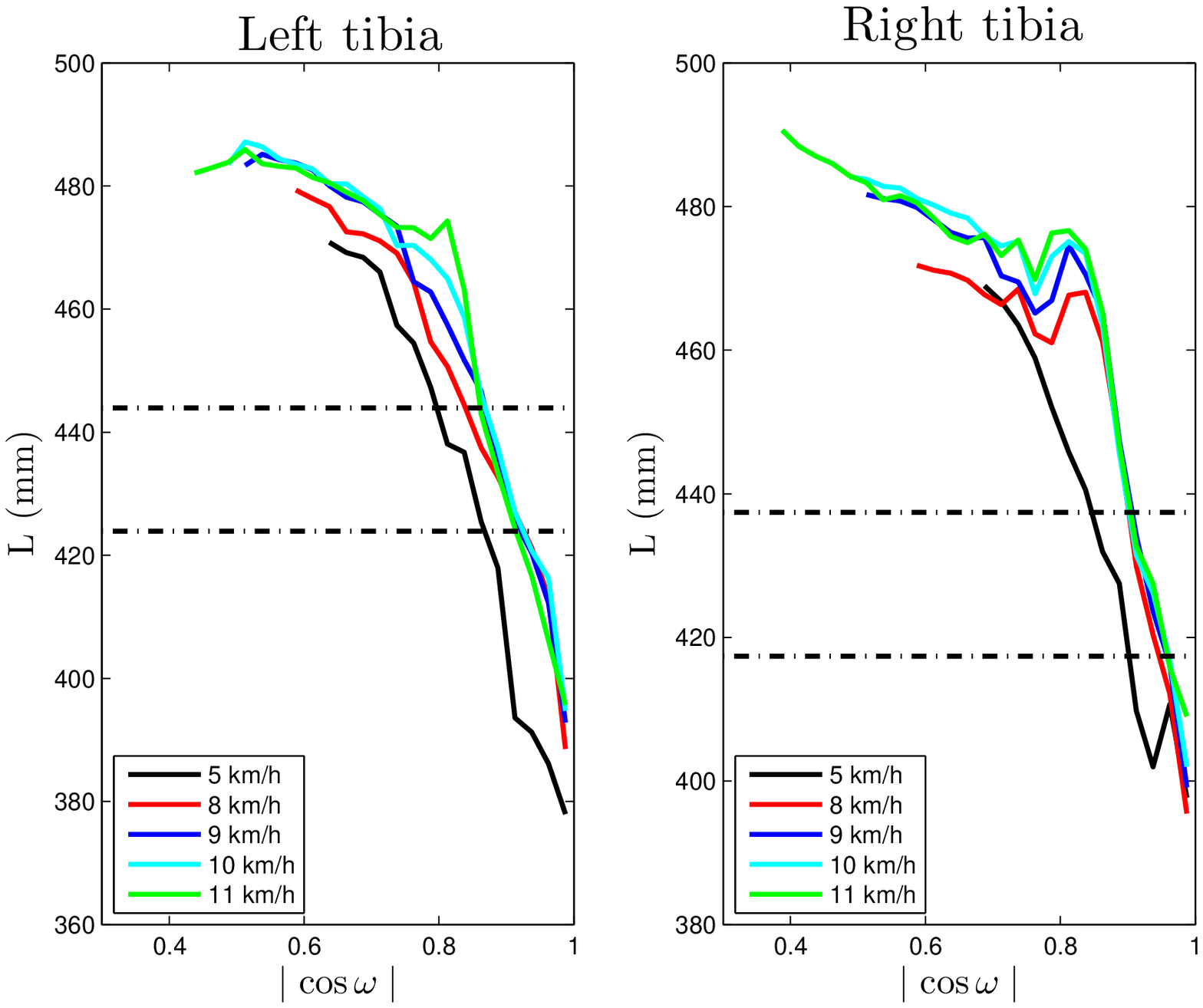}
%\vspace{-6cm}
\caption{\label{fig:LengthTibiaProfV1}The profile histogram of the tibial-bone length (separately for the left- and right-side bones) in $\lvert \cos\omega \rvert$ bins; the data has been captured with the original Kinect sensor.}
\end{center}
\end{figure}

\begin{figure}
\begin{center}
\includegraphics [width=15.5cm] {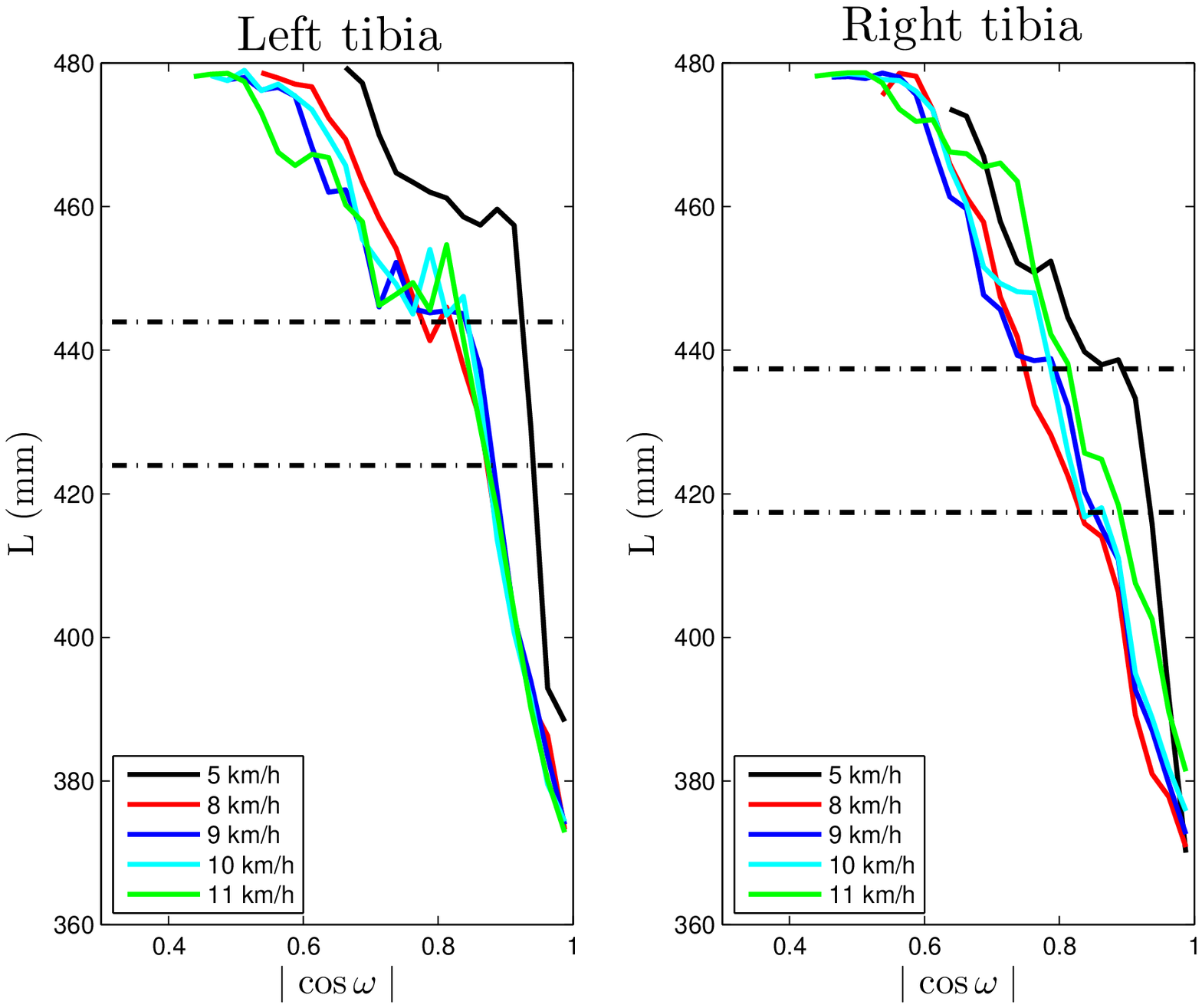}
%\vspace{-6cm}
\caption{\label{fig:LengthTibiaProfV2}The profile histogram of the tibial-bone length (separately for the left- and right-side bones) in $\lvert \cos\omega \rvert$ bins; the data has been captured with the upgraded Kinect sensor.}
\end{center}
\end{figure}

The analysis of the arm-bone lengths in terms of the inclination angle $\omega$ demonstrated that the results obtained with both sensors do not agree well with those of the static measurements; agreement occurs only in the 
walking-motion data for $\lvert \cos\omega \rvert \to 1$, where the bones are viewed at almost right angle by the sensor. We advanced an argument providing an explanation of this effect: it is mainly due to the systematic 
misplacement of the elbow nodes in the running motion, where this joint is kept in flexed position throughout the gait cycle. Regarding the leg bones, there is no doubt that the original Kinect sensor seriously overestimates 
the femoral-bone length. Both sensors overestimate the tibial-bone lengths for small viewing angles and underestimate them for large; discrepancies in the evaluated length of these bones lie in the vicinity of $9$-$11$ cm, see 
Figs.~\ref{fig:LengthTibiaProfV1} and \ref{fig:LengthTibiaProfV2}.

The present work corroborates earlier results \cite{pwbn}, obtained with the original sensor (the upgraded sensor was not available at the time that study was conducted), that Kinect cannot be easily employed in 
medical/health-relating applications requiring high accuracy. In most cases, the results obtained with the two sensors disagree with the static measurements and show a large range of variation within the gait cycle. The 
determination and application of corrections, needed in order to suppress these artefacts, comprises an interesting research subject.

Finally, we would like to comment on one misconception in the field of Medical Physics, namely that the results of studies using one subject, or only a few subjects, are not reliable. In our opinion, there are studies which call 
for statistics and studies in which statistics is superfluous. When comparing two measurement systems and the results (for a few subjects) come out sufficiently close, it does make sense to `pursue statistics' and obtain reliable 
estimates of averages, standard deviations, and ranges. On the contrary, there are occasions (in particular, in the validation of measurement systems) where serious discrepancies are found in the output already obtained from the 
first subject. Unless one explains why the comparison of the output of the two measurement systems failed for that one subject, one subject is sufficient in invalidating the application! Even in case that the tested measurement 
system failed for the specific subject (and that it does not fail for the general subject), its validation must be performed for every future subject separately, to guarantee the validity of the output \emph{on a case-by-case 
basis}; of course, this is not the essence of validations of measurement systems.

\begin{figure}
\begin{center}
\includegraphics [width=15.5cm] {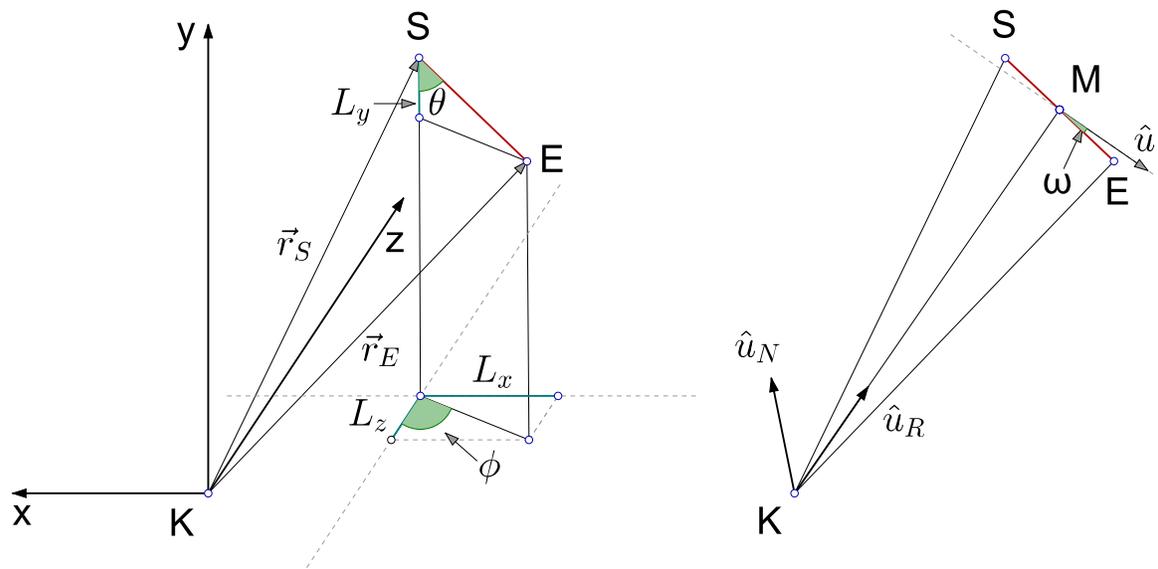}
%\vspace{-6cm}
\caption{\label{fig:CS}The coordinate system of the Kinect sensor. The endpoints of the specific bone are identified as S and E. The angles $\theta$ and $\phi$ define the orientation of the bone in space, 
whereas the angle $\omega$ pertains to the Kinect view of the bone; when Kinect views the bone at right angle, $\omega=0^\circ$. The (focal point of the) camera of the sensor appears in the figure as point K.}
\end{center}
\end{figure}

\begin{ack}
The original idea of investigating the subject of the present study belongs to S.~Roth and M.~Regniet, who (along with R.~Jain) had conducted a similar analysis of data obtained with the original Kinect sensor.

We are indebted to R.~Guntersweiler and L.~Lombriser for their support in the measurements obtained with the AICON system.

Fig.~\ref{fig:CS} has been produced with CaRMetal, a dynamic geometry free software (GNU-GPL license), first developed by R.~Grothmann and recently under E.~Hakenholz \cite{CMl}.\\
\end{ack}

{\bf Conflict of interest statement}

The authors certify that, regarding the material of the present paper, they have no affiliations with or involvement in any organisation or entity with financial or non-financial interest.

\newpage
\begin{table}[h!]
{\bf \caption{\label{tab:aicon}}}The values of the subject's bone lengths (in mm), obtained from a marker-based system (see Section \ref{sec:Results}). To account for incorrect placement of the markers, an uncertainty of $10$ mm 
is assumed in all cases, save for the femoral-bone lengths, where an overall uncertainty of $20$ mm is applicable (linear combination of the placement uncertainty of $10$ mm and of a $10$-mm uncertainty representing the systematic 
effects of Ref.~\cite{dotg}, as discussed in Section 2.2 of Ref.~\cite{mmr}). These values have been verified with a non-stretchable tape measure.
\vspace{0.6cm}
\begin{center}
\begin{tabular}{|c|c|}
\hline
Left humerus & $343$ \\
Right humerus & $344$ \\
Left ulna & $258$ \\
Right ulna & $271$ \\
Left femur & $434$ \\
Right femur & $436$ \\
Left tibia & $434$ \\
Right tibia & $427$ \\
\hline
\end{tabular}
\end{center}
\end{table}

\newpage
\begin{table}[h!]
{\bf \caption{\label{tab:Lengths}}}The average values of the subject's bone lengths (in mm), separately for the two Kinect sensors.
\vspace{0.6cm}
\begin{center}
\begin{tabular}{|c|c|c|c|c|c|}
\hline
 & $5$ km/h & $8$ km/h & $9$ km/h & $10$ km/h & $11$ km/h \\
\hline
\multicolumn{6}{|c|}{Original Kinect sensor} \\
\hline
Left humerus & $278(23)$ & $249(22)$ & $245(16)$ & $249(16)$ & $250(14)$ \\
Right humerus & $284.1(6.1)$ & $239(12)$ & $240(13)$ & $244(18)$ & $247(15)$ \\
Left ulna & $293(18)$ & $271(27)$ & $265(29)$ & $253(24)$ & $251(26)$ \\
Right ulna & $277(15)$ & $282(24)$ & $268(24)$ & $249(30)$ & $251(30)$ \\
Left femur & $531(29)$ & $517(25)$ & $513(30)$ & $510(28)$ & $503(30)$ \\
Right femur & $519(23)$ & $512(17)$ & $512(25)$ & $508(25)$ & $506(30)$ \\
Left tibia & $403(36)$ & $431(35)$ & $436(36)$ & $440(38)$ & $443(40)$ \\
Right tibia & $418(30)$ & $439(35)$ & $445(37)$ & $447(37)$ & $451(36)$ \\
\hline
\multicolumn{6}{|c|}{Upgraded Kinect sensor} \\
\hline
Left humerus & $268(20)$ & $241(10)$ & $240.3(6.6)$ & $243.3(7.4)$ & $244.0(8.0)$ \\
Right humerus & $273(10)$ & $236(12)$ & $235(13)$ & $238(14)$ & $239(14)$ \\
Left ulna & $248(16)$ & $253(30)$ & $255(30)$ & $244(28)$ & $241(27)$ \\
Right ulna & $253(11)$ & $259(33)$ & $246(30)$ & $228(21)$ & $230(23)$ \\
Left femur & $409(42)$ & $422(33)$ & $428(39)$ & $423(41)$ & $419(40)$ \\
Right femur & $427(35)$ & $432(28)$ & $438(34)$ & $430(38)$ & $418(41)$ \\
Left tibia & $422(48)$ & $420(41)$ & $422(42)$ & $423(44)$ & $425(45)$ \\
Right tibia & $409(48)$ & $411(39)$ & $415(40)$ & $418(41)$ & $428(42)$ \\
\hline
\end{tabular}
\end{center}
\end{table}

\clearpage
% ============= FIGURE 1

\clearpage
% ============= FIGURE 2

\clearpage
% ============= FIGURE 3

\clearpage
% ============= FIGURE 4

\clearpage
% ============= FIGURE 5

\clearpage
% ============= FIGURE 6

\clearpage
% ============= FIGURE 7

\clearpage
% ============= FIGURE 8

\clearpage
% ============= FIGURE 9

\end{document}